# How Much Does Auger Spectroscopy Add to XAS? A Fisher-information Analysis of Core-Hole Dynamics in $Ar^+$

**Hadeel Hamid Saud**

Department of Physics, Istanbul Kültür University, Istanbul, Türkiye

Corresponding author: hadeel86saud@gmail.com

ORCID: 0009-0005-2365-3885

## Abstract

X-ray absorption spectroscopy (XAS) and resonant-Auger spectroscopy probe different stages of core-hole excitation and decay, but distinct spectra need not provide independent information about decay parameters. We quantify their complementarity for the $Ar^+$ $2p \rightarrow 3d$ manifold using state-resolved atomic calculations and a two-parameter Fisher-information model. A common scale of primary autoionization rates $s_\Gamma$ and a selective spectator-$pp$ rescaling $\eta_{pp}$ both alter XAS linewidths. Their spectral responses are nearly collinear because the spectator-$pp$ fraction varies weakly among populated intermediate states. We compare XAS with an effective, scan-integrated, normalized Auger spectrum defined on a final-state-energy coordinate, with fixed excitation weights and final-state widths. Within this reduced observable, the common rate scale cancels from the normalized branching fractions, while $\eta_{pp}$ redistributes intensity among ionic final states. This cancellation is a property of the adopted model, not of photon-energy-resolved resonant-Auger spectroscopy generally. Under the reference count and resolution assumptions, XAS alone yields a Cramér−Rao lower-bound (CRLB) corresponding to a 1.20% local fractional standard uncertainty in the excitation-weighted total natural width; adding Auger reduces this uncertainty by only 0.70%. In contrast, the uncertainty in the spectator-$pp$ branching fraction falls by 99.19%, with similarly large gains for partial-rate scales. A generalized Fisher comparison gives a 123.5-fold improvement on the standard-deviation scale along the most enhanced information mode. Thus, the principal benefit of the complementary Auger observable is to resolve a nearly degenerate decay-channel decomposition rather than substantially refine the total natural width.

**Keywords:** X-ray absorption spectroscopy (XAS); Resonant-Auger spectroscopy; Core-hole dynamics; Fisher information; Parameter identifiability; Multichannel autoionization.

# 1. Introduction

Core-level excitation creates a transient electronic state whose decay dynamics are encoded only indirectly in spectroscopic observables. The finite lifetime of a core-excited state contributes to the intrinsic Lorentzian broadening of its spectral profile, while the experimentally observed line shape is further shaped by instrumental resolution, overlapping transitions, and counting statistics. Extracting decay dynamics from such spectra is therefore an inverse problem: the relevant question is not only whether a spectral feature is sensitive to a lifetime or decay rate, but which combinations of the underlying physical parameters can be identified from the measurement and with what precision. The separation of intrinsic lifetime broadening from instrumental contributions has long been recognized as a central problem in core-level spectroscopy [1,2].

X-ray absorption spectroscopy (XAS) and Auger electron spectroscopy interrogate different stages of the same excitation−decay sequence. XAS probes the energies, oscillator strengths, and lifetime-broadened structure of the intermediate core-excited manifold. Auger spectroscopy probes the subsequent redistribution of population among ionic final states and therefore provides access to the channel structure of the decay. Experimental studies of argon have demonstrated both lifetime sensitivity in high-resolution core-level spectra [3] and pronounced channel selectivity in resonant-Auger decay [4,5], while multielectron measurements have further shown how intermediate and final ionic states can be identified within complex relaxation cascades [6,7].

Spectral complementarity, however, does not necessarily imply information complementarity. Two observables may look substantially different while constraining nearly the same combination of physical parameters. Conversely, a second observable may add little information about a total decay width while supplying precisely the information required to determine how that width is partitioned among competing decay channels. This distinction becomes especially important in multichannel core-hole relaxation, where an accurately determined lifetime-like quantity does not by itself imply that the underlying partial rates or branching fractions are separately identifiable.

The relevant question is therefore not simply whether XAS and Auger spectra are spectroscopically distinct, but whether they constrain different directions of the same inverse problem.

Fisher information provides a natural framework for making this distinction quantitative. For a specified statistical measurement model, the Fisher-information matrix characterizes the local sensitivity of the probability distribution to changes in the physical parameters, while the Cramér−Rao lower bound (CRLB) provides the corresponding local precision bound for unbiased estimators [8,9]. The matrix structure is essential: parameter-by-parameter sensitivities alone do not reveal whether two spectral responses are nearly redundant. Nearly collinear derivatives generate strong parameter correlations and small Fisher eigenvalues, exposing weakly identifiable combinations even when the spectrum is individually sensitive to each parameter [10–12]. Fisher-information methods have been applied previously to spectroscopic parameter estimation, including broadened line-shape analysis and quantum-enhanced spectral measurements [13,14]. The use of complementary measurements to resolve parameter degeneracies is well established in other fields, notably in Fisher-matrix analyses of combined cosmological data sets [15], and in studies of sloppy multiparameter models [11,12]; here, this viewpoint is applied to core-hole spectroscopy.

This viewpoint suggests a specific mechanism by which XAS and Auger spectroscopy can become complementary. XAS may constrain a lifetime-broadened combination of several decay rates with high precision while remaining unable to determine how the total width is decomposed into its individual contributions. A channel-selective Auger observable may then contain little or no direct information about the common decay-rate scale, yet still yield a large improvement in the joint inference by constraining the correlated channel-decomposition degree of freedom. In such a case, the value of the second measurement arises from degeneracy breaking rather than from two independent measurements of the same lifetime quantity.

$Ar^+$ provides a state-resolved system in which this mechanism can be examined explicitly. The L-shell photoabsorption of $Ar^+$ has been measured with photon−ion merged-beam techniques [16,17], and the spectroscopy, lifetimes, and Auger decay pathways of a $2p$ inner-shell hole in $Ar^+$ have been studied by multielectron coincidence spectroscopy [18]. The $2p \rightarrow 3d$ core-excited manifold is modeled here using atomic-structure, radiative-transition, and autoionization calculations performed with the Flexible Atomic Code (FAC) [19]. The calculation contains 65 core-excited fine-structure levels populated by 76 E1 transitions, 7397 primary autoionization channels, and 115 $Ar^{2+}$ final levels distributed among five configuration groups. XAS and Auger forward models are therefore constructed from a common microscopic description of the same excitation and decay manifold rather than from independent phenomenological spectra.

The inverse problem is parameterized by two dimensionless quantities with distinct physical roles. The parameter $s_\Gamma$ applies a common multiplicative scale to the primary autoionization rates, whereas $\eta_{pp}$ selectively rescales the spectator-$pp$ contribution. Both quantities modify the intermediate–state natural widths entering XAS. Their spectral effects become difficult to distinguish, however, when the spectator-$pp$ fraction varies only weakly across the populated intermediate states. In the limiting case of an identical spectator-$pp$ fraction for every state, the local XAS responses to $s_\Gamma$ and $\eta_{pp}$ become proportional and the corresponding two-parameter Fisher matrix becomes rank one.

The normalized final-state-resolved Auger observable has a qualitatively different information structure. Primary autoionization rates are converted into branching fractions and aggregated into an effective, incoherent, excitation-weighted final-state spectrum. Within this construction, the common scale $s_\Gamma$ cancels exactly from the normalized branching fractions, whereas $\eta_{pp}$ changes their relative distribution among ionic final states. Auger therefore carries no direct Fisher information on $s_\Gamma$, but remains strongly sensitive to the channel-selective parameter $\eta_{pp}$. The adopted rate-based model represents a narrow-band final-state-resolved resonant-Auger measurement and is not intended as a fully coherent Kramers–Heisenberg description of resonant-Auger scattering [20].

The resulting information geometry is strongly anisotropic. Under the reference count and resolution model, XAS alone constrains the excitation-weighted total natural width to a local fractional CRLB of approximately 1.20%, even though the two-parameter XAS Fisher matrix is almost singular, with a correlation coefficient close to $-1$. Adding Auger reduces the standard uncertainty of the total natural width by only 0.70%. The gain in channel-resolved quantities is much larger: the CRLB standard uncertainty is reduced by 99.19% for the excitation-weighted spectator-$pp$ branching fraction, 97.82% for the non-$pp$ rate scale, and 96.25% for the $pp$ partial-rate scale. The principal contribution of Auger is therefore not a substantially improved estimate of an already well-constrained total width, but reconstruction of the decay-channel decomposition hidden within it.

This anisotropy is quantified by comparing the joint and XAS Fisher matrices through a generalized eigenvalue problem. Because the Auger contribution is a rank-one update, the relative-information eigenvalues are exactly 1 and approximately $1.53 \times 10^4$. Thus, one generalized-information mode is left unchanged, whereas the complementary mode is enhanced by more than four orders of magnitude in Fisher information, corresponding to a factor of approximately 123.5 on the associated standard-deviation scale. The linear parameter combination whose local CRLB is left unchanged closely follows the total-width functional. This near coincidence provides a geometric explanation for why the total width benefits only weakly from Auger information while channel-resolved quantities improve dramatically.

The robustness and measurement implications of this information structure are examined at several levels. Finite-difference tests establish the numerical stability of the Fisher values, while an auxiliary Auger representation and explicit Poisson reconstruction test the local Fisher approximation. Instrumental nuisance analysis shows that uncertainty in the XAS resolution primarily affects inference of the absolute decay scale, whereas the Auger-driven determination of the channel-selective parameter remains essentially unchanged over the tested range. A fixed detected-count analysis further shows that information necessity does not imply count dominance: when $s_\Gamma$ is the target quantity, the optimal allocation assigns only about 26% of the expected detected events to Auger spectroscopy, while standard multiparameter design criteria select different interior allocations. These results concern detected-count allocation rather than beam time or experimental cost.

The objective of this work is therefore not merely to demonstrate that adding an independent observable increases Fisher information. Instead, we address a more specific physical question: how much non-redundant information does resonant-Auger spectroscopy add to XAS, which aspects of multichannel core-hole decay does that information constrain, and when is the additional observable required to resolve otherwise degenerate decay dynamics?

By separating information on the total decay scale from information on its channel decomposition, and by distinguishing direct sensitivity from degeneracy breaking, the present work connects state-resolved atomic decay calculations with quantitative identifiability analysis. More broadly, it shows that the value of a complementary spectroscopic observable is determined not simply by how different its spectrum appears, but by whether it supplies an information direction that is absent or poorly constrained in the original measurement.

# 2. Theoretical and Computational Methods

## 2.1. Atomic-Structure and Decay Model of Ar+

Atomic-structure and decay calculations were performed for singly ionized argon, $Ar^{+}$, using the Flexible Atomic Code (FAC) [19]. The closed $1s^2 2s^2$ subshells are omitted from the configuration notation below. The ground configuration was represented as $2p^6 3s^2 3p^5$, and the resonant core-excited manifold was generated by the $2p \rightarrow 3d$ excitation, $2p^5 3s^2 3p^5 3d$. The FAC structure calculation produced 67 fine-structure levels: the two components of the $Ar^{+}$ ground term, $J = 3/2$ and $J = 1/2$, together with 65 core-excited levels spanning 251.679–260.497 eV on the FAC energy scale. The excitation dataset contained 76 electric-dipole (E1) transitions from the two initial fine-structure levels, reaching 47 distinct core-excited levels. The remaining 18 calculated levels carry zero E1 excitation weight within the adopted model. The E1 angular-momentum selection rules restrict the accessible upper levels to $J = 1/2, 3/2,$ or $5/2$, subject also to the required parity change.

The initial populations were assumed to follow statistical weights proportional to $2J + 1$, giving

$$P_{3/2} = \frac{2}{3}, \qquad P_{1/2} = \frac{1}{3}.$$

To distinguish individual excitation transitions from intermediate levels, $g$ denotes an initial fine-structure level and $i$ a core-excited level. The normalized excitation-transition weight was defined as

$$w_{gi} = \frac{P_g f_{gi}}{\sum_{g',i'} P_{g'} f_{g'i'}},$$

where $f_{gi}$ is the FAC absorption oscillator strength. The corresponding intermediate-state excitation weight was obtained by summing over the initial levels,

$$w_i^{exc} = \sum_g w_{gi}, \qquad \sum_i w_i^{exc} = 1.$$

The transition weights $w_{gi}$ enter the XAS spectral construction, whereas $w_i^{exc}$ is used for excitation-weighted averages of state-resolved decay quantities. Primary autoionization from the $2p^5 3s^2 3p^5 3d$ manifold was evaluated state by state. Five $Ar^{2+}$ final-configuration groups were included. Participator decay was represented by $2p^6 3s^2 3p^4$ and $2p^6 3s 3p^5$, whereas spectator decay was represented by $2p^6 3s^2 3p^3 3d$, $2p^6 3s 3p^4 3d$, $2p^6 3p^5 3d$. These five configuration groups generated 115 $Ar^{2+}$ fine-structure levels, and the primary-autoionization dataset contained 7397 resolved decay channels. For each intermediate level $i$, the total primary Auger rate was calculated by summing the partial rates $A_{if}$ over the energetically accessible final states included in the model,

$$A_i^{Auger} = \sum_f A_{if}$$

The corresponding Auger-only lifetime and natural full width at half maximum (FWHM) were defined as

$$\tau_i^{Auger} = \frac{1}{A_i^{Auger}}, \quad \Gamma_i^{Auger} = \hbar A_i^{Auger}.$$

The Lorentzian half width at half maximum (HWHM) is

$$\gamma_i^{Auger} = \frac{\Gamma_i^{Auger}}{2}.$$

Radiative competition was estimated from the calculated E1 decay channels returning to the two initial ground-term fine-structure components. For each intermediate state,

$$A_i^{rad} = \sum_g A_{i\to g}^{E1},$$

where the sum includes only these explicitly calculated radiative channels. The total decay rate and corresponding natural width within the adopted model were

$$A_i^{tot} = A_i^{Auger} + A_i^{rad},$$

$$\Gamma_i^{tot} = \hbar A_i^{tot}.$$

The excitation-weighted radiative branching fraction was defined as

$$\bar{b}_{rad} = \sum_i w_i^{exc} \frac{A_i^{rad}}{A_i^{tot}},$$

with the corresponding Auger fraction

$$\bar{b}_{Auger} = \sum_i w_i^{exc} \frac{A_i^{Auger}}{A_i^{tot}} = 1 - \bar{b}_{rad}.$$

For the explicitly included radiative channel set, these fractions are $\bar{b}_{red} = 3.38 \times 10^{-4}$, $\bar{b}_{Auger} = 0.999662$.

The excitation-weighted mean natural widths were

$$\bar{\Gamma}_{Auger} = \sum_i w_i^{exc} \Gamma_i^{Auger} = 0.098268 \text{ eV},$$

and

$$\bar{\Gamma}_{tot} = \sum_i w_i^{exc} \Gamma_i^{tot} = 0.098296 \text{ eV}.$$

These quantities are excitation-weighted averages of the individual state widths, rather than the FWHM of the composite absorption spectrum. The very small radiative contribution shows that autoionization dominates the decay within the explicitly included channel space. The calculated radiative rates are nevertheless retained as fixed contributions in the parameterized XAS model described in Sec. 2.3.

The final-state lifetime contribution to the Auger line shapes was evaluated separately. Secondary autoionization from the 115-level $Ar^{2+}$ manifold into the included $Ar^{3+}$ configurations was calculated with FAC. Forty-six $Ar^{2+}$ levels have nonzero secondary-autoionization rates. For such a final level $f$, the natural width and Lorentzian HWHM were defined as

$$\Gamma_f = \hbar A_f^{sec}, \gamma_f \frac{\Gamma_f}{2},$$

where $A_f^{sec}$ is the total rate summed over the included secondary-autoionization channels. Final states without an accessible secondary-autoionization channel within this configuration space were assigned $\Gamma_f = 0$ for this contribution. The nonzero calculated final-state natural FWHM values span $1.65 \times 10^{-5}$ eV $\leq \Gamma_f \leq 0.897$ eV.

Primary transitions terminating in these autoionizing $Ar^{2+}$ final states account for approximately 23.17% of the total excitation-weighted primary Auger decay strength. The resulting state-specific final-state widths are used in the effective final-state-resolved resonant-Auger spectral model described in Sec. 2.2.

The spectral construction uses the FAC rates to combine excitation and decay pathways incoherently. The calculation provides state-resolved level energies, oscillator strengths, primary-autoionization rates, radiative rates, and secondary-autoionization rates. However, the rate-based forward model does not include the relative complex phases or interference between competing excitation−decay amplitudes. It is therefore an effective, incoherent description rather than a coherent Kramers−Heisenberg treatment of resonant-Auger scattering [20]. This distinction defines the scope of the present atomic model and is carried explicitly into the spectral construction described in Sec. 2.2.

## 2.2. XAS and Resonant-Auger Forward Models

The state−resolved atomic data described in Sec. 2.1 were used to construct forward models for the X-ray absorption and resonant-Auger observables. Both spectra originate from the same $2p \rightarrow 3d$ core-excited manifold, but they probe different stages of the excitation−decay sequence.

For XAS, each electric-dipole transition from an initial ground-state fine-structure component $g$ to a core-excited level $i$ was represented by a Voigt profile centered at the corresponding FAC transition energy $E_{gi} = E_i - E_g$, where $E_i$ and $E_g$ are the energies of the intermediate and initial states, respectively. The normalized excitation-transition weights $w_{gi}$ and the corresponding intermediate-state weights $w_i^{exc}$ were defined in Sec. 2.1 from the initial-state statistical populations and FAC oscillator strengths. These weights were held fixed when the primary decay parameters were varied. An additional common strength rescaling present in the archived XAS transition table cancels upon normalization of the spectral shape and therefore does not affect the fixed-expected-count Fisher analysis.

The Lorentzian component of each displayed reference XAS resonance was determined from the state-specific primary autoionization rate of the corresponding intermediate level,

$$\Gamma_i^{Auger} = \hbar A_i^{Auger}, \qquad \gamma_i^{Auger} = \frac{\Gamma_i^{Auger}}{2}.$$

The independently calculated E1 radiative contribution within the included channel set is very small. Including it changes the excitation-weighted mean of the individual natural FWHMs from 0.098268 eV to 0.098296 eV, a difference of 2.81 × 10−5 eV. The values represent averages of state-specific natural widths, rather than the FWHM of the composite absorption spectrum. The Auger-only widths therefore provide an adequate representation of the displayed reference natural broadening. In the parameterized inference model introduced in Sec. 2.3, the calculated radiative contribution is retained as a fixed term while the primary autoionization rates are varied.

Instrumental broadening of the XAS spectrum was represented by a Gaussian response with FWHM

$$\Delta E_{XAS}^{G} = 0.140 \text{ eV},$$

corresponding to

$$\sigma_{XAS} = \frac{0.140}{2\sqrt{2In\,2}} \simeq 0.05945 \text{ eV}.$$

The reference XAS spectrum was generated over $250 \ \leq E \ \leq 262$ eV. An 8000-point grid was used for the displayed reference spectrum, whereas the Fisher-information calculations used a uniform 4000-point grid over the same interval, corresponding to an energy spacing of approximately 3 meV. The XAS forward model is

$$S_{XAS}(E) = \sum_{g,i} w_{gi}\, V\left(E - E_{gi};\ \sigma_{XAS}, \gamma_i\right),$$

where $V(x;\ \sigma, \gamma)$ denotes a normalized Voigt profile with energy offset $x$, Gaussian standard deviation $\sigma$, and Lorentzian HWHM $\gamma$. Retaining the transition indices $(g, i)$ preserves the distinct excitation energies associated with the two initial fine-structure components. The parameter dependence of $\gamma_i$ used in the inference analysis is defined in Sec. 2.3.

The resonant-Auger observable was constructed from the same core-excited manifold as an effective, incoherent, excitation-weighted sum over the state-resolved primary autoionization channels. For an intermediate state $i$ decaying into an $Ar^{2+}$ final state $f$, the reference branching fraction is

$$B_{if} = \frac{A_{if}}{\sum_{f'} A_{if'}},$$

where $A_{if}$ is the FAC partial autoionization rate. Contributions terminating in the same $Ar^{2+}$ final level were combined to define the final-state weight

$$W_f = \sum_{i} w_i^{exc} B_{if}.$$

The fixed excitation weights represent an idealized scan-integrated average corresponding to uniform spectral photon fluence over the included resonances. Given the small calculated radiative fraction reported in Sec. 2.1, these Auger event weights adopt a unit non-radiative-yield approximation: radiative competition is retained in the XAS lifetime widths but omitted from the normalized Auger branching weights. This approximation is distinct from the subsequent normalization of the electron spectrum within its finite analysis window.

The resulting observable was represented on an effective final-state-energy coordinate $u$, in which contributions from the two initial fine-structure components are aggregated at the calculated ionic final-state energies,

$$S_{Auger}(u) = \sum_{f} W_f V\left(u - E_f;\, \sigma_A, \gamma_f\right),$$

where the sum runs over the 115 calculated $Ar^{2+}$ final levels, $E_f$ is the corresponding final-state energy, and $\sigma_A = 0.12\ eV$ is the adopted Gaussian standard deviation of the electron-analyzer response.

This effective final-state-energy coordinate should be distinguished from a fully initial-state-resolved representation. For an initial component $g$, energy conservation gives a binding-energy line position of the form

$$E_{b,gf} = E_f - E_g + C,$$

where C denotes a fixed energy-reference offset. Because the two initial fine-structure components have different energies, their contributions generally cannot be represented by a single common threshold offset. The present construction instead combines contributions at the same calculated final-state energy $E_f$. A known common translation of the effective energy coordinate leaves the Fisher information unchanged when the analysis window is translated consistently.

For a final state stable against secondary autoionization within the included configuration space, $\Gamma_f = 0$ , $\gamma_f = 0$. For an unstable final state,

$$\Gamma_f = \hbar A_f^{sec},\ \ \gamma_f = \frac{\Gamma_f}{2},$$

where $A_f^{sec}$ is the total calculated secondary-autoionization rate for $Ar^{2+} \rightarrow Ar^{3+} + e^-$.

The final-state-resolved Auger spectrum was evaluated on a uniform 16000-point grid extending 5 eV beyond the lowest and highest calculated final-state energies. The final-state energies and Lorentzian widths were held fixed throughout the two-parameter inference analysis.

This construction is interpreted as an effective, scan-integrated representation motivated by the narrow-band resonant-Auger Raman regime. The photon energy is conceptually scanned across the $2p \rightarrow 3d$ manifold, while the integrated excitation contributions are represented by fixed weights rather than by an explicit coherent photon-energy-dependent scattering calculation. The emitted-electron observable is analyzed on the effective final-state-energy coordinate defined above. Under idealized resonant-Raman conditions, where the excitation bandwidth is narrower than the natural width of the intermediate core-excited state, the natural Lorentzian contribution to the emitted-electron line shape can be governed by the lifetime of the ionic final state rather than directly by the intermediate-state width [5,21]. In the present construction, the excitation weights $w_i^{exc}$ are defined from the initial-state populations and FAC oscillator strengths and are therefore independent of $\Gamma_i$. Together with the fixed final-state widths and the adopted unit non-radiative-yield approximation, this ensures that the intermediate-state widths do not enter $S_{Auger}$.

Outside this narrow-band resonant-Raman regime, intermediate-state lifetime broadening can re-enter the emitted-electron line shape. In the normal two-step Auger limit, the natural transition width contains contributions from both the intermediate- and final-state widths, $\Gamma_i + \Gamma_f$, so the Auger observable can acquire direct sensitivity to the intermediate-state decay scale. The present model does not include coherent interference between overlapping intermediate resonances and should therefore not be interpreted as a full Kramers−Heisenberg treatment of resonant-Auger scattering [20].

For the statistical inference, both spectra were normalized to fixed expected detected-count budgets, with spectral-bin counts modeled as independent Poisson variables. The expected counts in spectral bin $k$ were

$$\mu_k^{XAS} = N_{XAS} \frac{S_{XAS}(E_k)}{\sum_j S_{XAS}(E_j)}, \quad N_{XAS} = 10^5,$$

and

$$\mu_k^{Auger} = N_{Auger} \frac{S_{Auger}(u_k)}{\sum_j S_{Auger}(u_j)}, \quad N_{Auger} = 8.0 \times 10^4.$$

Consequently,

$$\sum_k \mu_k^{XAS} = N_{XAS}, \quad \sum_k \mu_k^{Auger} = N_{Auger}.$$

This fixed-expected-count normalization deliberately isolates information carried by spectral shape and relative channel weights from information associated with the absolute photon or electron yield. The expected count totals are held fixed, while the realized Poisson counts may fluctuate between measurements. No artificial phase assignments were introduced because the rate-based FAC data used here do not provide the relative complex phases required for coherent sums over competing excitation−decay amplitudes. The sensitivity to finite excitation bandwidth within the adopted effective line-shape framework is examined separately in Sec. 2.4.

## 2.3. Parameterization, Fisher Information, and Identifiability

To quantify the complementary information carried by XAS and the final-state-resolved Auger observable, the state-resolved atomic model was parameterized by two dimensionless scaling parameters, $\theta = (s_\Gamma, \eta_{pp})$, with the reference FAC calculation corresponding to $s_\Gamma = \eta_{pp} = 1$.

The parameter $s_\Gamma$ applies a common multiplicative scale to all primary autoionization rates. It therefore controls the overall scale of the primary non-radiative decay, but it should not be identified with the total intermediate-state width when $\eta_{pp}$ is allowed to vary independently. The parameter $\eta_{pp}$, in contrast, selectively rescales the spectator-$pp$ contribution and therefore changes the channel decomposition of the primary Auger decay.

Starting from the reference FAC partial autoionization rates $A_{if}^{(0)}$, the parameterized rates were defined as

$$A_{if}(\theta) = \begin{cases} s_\Gamma \eta_{pp} A_{if}^{(0)}, & f \in \mathcal{P}_{pp}, \\ s_\Gamma A_{if}^{(0)}, & f \notin \mathcal{P}_{pp}, \end{cases}$$

where $\mathcal{P}_{pp}$ denotes the spectator-$pp$ final-state group. For each intermediate state $i$, the total primary autoionization rate is therefore

$$A_i^{Auger}(\theta) = s_\Gamma\left[\eta_{pp}A_{i,pp}^{(0)} + A_{i,rest}^{(0)}\right].$$

For the XAS model, the total decay rate entering the intermediate-state natural linewidth was written as $A_i^{XAS}(\theta) = A_i^{Auger}(\theta) + A_i^{rad}$, where $A_i^{rad}$ was held fixed at its calculated reference value. As discussed in Sec. 2.2, the explicitly included radiative contribution produces only a very small correction to the reference linewidth scale. Retaining it as a fixed term nevertheless preserves the distinction between the varied primary autoionization rates and the independently calculated radiative channels. The corresponding Lorentzian HWHM entering the XAS forward model is

$$\gamma_i(\theta) = \frac{\hbar}{2}A_i^{XAS}(\theta).$$

The two parameters affect the XAS and Auger observables differently. Both $s_\Gamma$ and $\eta_{pp}$ modify the state-dependent intermediate widths entering XAS. In the normalized final-state-resolved Auger observable, however, the branching fraction for the decay $i \to f$ is

$$B_{if}(\theta) = \frac{A_{if}(\theta)}{\sum_{f'} A_{if'}(\theta)}.$$

Because $s_\Gamma$ multiplies all primary autoionization channels originating from a given intermediate state, it cancels exactly between numerator and denominator. Consequently,

$$B_{if} = B_{if}(\eta_{pp}),$$

with no direct dependence on $s_\Gamma$. The Lorentzian widths assigned to the final-state-resolved Auger features are determined independently from the calculated $Ar^{2+}$ final-state lifetimes and are likewise not rescaled by $s_\Gamma$. Therefore, within the adopted normalized Auger model,

$$\frac{\partial S_{Auger}}{\partial s_\Gamma} = 0,$$

whereas

$$\frac{\partial S_{Auger}}{\partial \eta_{pp}} \neq 0.$$

This structural asymmetry is central to the inverse problem considered here: Auger carries no direct information on the common primary-rate scale, but it can strongly constrain the channel-selective parameter that is correlated with that scale in XAS.

For the statistical analysis, the spectral-bin counts were modeled as independent Poisson variables,

$$n_k \sim \text{Poisson}\,[\mu_k(\theta)].$$

For an observable with expected bin counts $\mu_k(\theta)$, the Poisson Fisher-information matrix [8,13] was evaluated as

$$F_{\alpha\beta} = \sum_k \frac{1}{\mu_k(\theta)} \frac{\partial \mu_k}{\partial \theta_\alpha} \frac{\partial \mu_k}{\partial \theta_\beta}$$

For the two-parameter model, the Fisher matrix was written in the ordered basis ($s_\Gamma$, $\eta_{pp}$) as

$$F(\theta) = \begin{pmatrix} F_{ss} & F_{s\eta} \\ F_{s\eta} & F_{\eta\eta} \end{pmatrix},$$

where

$$F_{ss} \equiv F_{s_\Gamma s_\Gamma}, \quad F_{s\eta} \equiv F_{s_\Gamma \eta_{pp}}, \qquad F_{\eta\eta} \equiv F_{\eta_{pp}\eta_{pp}}.$$

Unless otherwise specified, the reference expected detected-count budgets were

$$\mathrm{N_{XAs}} = 10^5, \quad \mathrm{N_{Auger}} = 8.0 \times 10^4.$$

Because the XAS and Auger measurements are treated as statistically independent, their Fisher matrices add,

$$\mathrm{F_{joint}} = \mathrm{F_{XAS}} + \mathrm{F_{Auger}}.$$

The purpose of this combination is not merely to exploit the additivity of independent Fisher information, but to determine whether the second observable constrains a parameter-space direction that remains weakly identifiable from the first.

For a nonsingular Fisher matrix, the local covariance of an unbiased estimator satisfies the Cramér–Rao inequality [8,9],

$$\mathrm{Cov}(\hat{\theta}) \succcurlyeq F^{-1}.$$

The marginalized CRLB standard uncertainty of parameter $\theta_\alpha$ was therefore defined as

$$\sigma_\alpha^{CRLB} = \sqrt{(F^{-1})_{\alpha\alpha}},$$

and the corresponding local parameter correlation as

$$\rho_{\alpha\beta} = \frac{(F^{-1})_{\alpha\beta}}{\sqrt{(F^{-1})_{\alpha\alpha}(F^{-1})_{\beta\beta}}}.$$

Local identifiability was characterized using the rank and eigenvalue spectrum of the Fisher matrix. For a nonsingular two-parameter Fisher matrix, the condition number was defined as

$$\kappa(F) = \frac{\lambda_{max}}{\lambda_{min}}.$$

A small $\lambda_{min}$ together with a large $\kappa$ indicates a highly elongated local uncertainty geometry and therefore a nearly degenerate parameter combination. Exact rank deficiency was interpreted as structural local non-identifiability and was not removed through arbitrary numerical regularization.

Because ordinary Fisher eigenvectors and condition numbers depend on the chosen parameter coordinates, they were used as diagnostics within the dimensionless ($s_\Gamma$, $\eta_{pp}$) parameterization rather than as coordinate-independent physical observables. The Fisher rank, by contrast, is preserved under nonsingular local reparameterization. To complement these coordinate-specific diagnostics, the covariance was therefore propagated to several physically interpretable derived quantities. The excitation-weighted total natural width was defined as

$$\bar{\Gamma}_{tot}(\theta) = \hbar \sum_i w_i^{exc} \left[ s_\Gamma \left( \eta_{pp} A_{i,pp}^{(0)} + A_{i,rest}^{(0)} \right) + A_i^{rad} \right].$$

The excitation-weighted spectator-$pp$ branching fraction was defined as

$$\bar{B}_{pp}(\eta_{pp}) = \sum_i w_i^{exc} \frac{\eta_{pp} A_{i,pp}^{(0)}}{\eta_{pp} A_{i,pp}^{(0)} + A_{i,rest}^{(0)}}.$$

Two additional partial-rate scale coordinates were introduced,

$s_{non-pp} = s_\Gamma$, and $s_{pp} = s_\Gamma \eta_{pp}$.

For any vector of derived quantities $\phi(\theta)$, the local covariance was obtained by Jacobian propagation,

$C_{\phi} = J C_{\theta} J^{T}$, where $C_{\theta} = F^{-1}$, $J_{\alpha\beta} = \frac{\partial \phi_{\alpha}}{\partial \theta_{\beta}}$.

This transformation was used to distinguish information about the total decay width from information about its decomposition into $pp$ and non-$pp$ contributions.

The fractional reduction in CRLB standard uncertainty produced by adding Auger spectroscopy was quantified, for a model parameter or derived quantity $\phi$, as

$$G_{\phi}^{(\sigma)} = 1 - \frac{\sigma_{\phi}^{joint}}{\sigma_{\phi}^{XAS}},$$

for quantities having a finite XAS-only uncertainty. The role of parameter correlation was examined further through the Schur complement. Treating $\eta_{pp}$ as a nuisance parameter, the effective Fisher information associated with $s_{\Gamma}$ is

$$F_{s_{\Gamma}}^{eff} = F_{ss} - \frac{F_{s\eta}^{2}}{F_{\eta\eta}},$$

with marginalized variance bound

$$\sigma_{s_{\Gamma}}^{2} \geq \frac{1}{F_{s_{\Gamma}}^{eff}}.$$

The second term represents the information penalty arising from correlation between the common primary-rate scale and the channel-selective parameter. This form therefore separates direct sensitivity to $s_{\Gamma}$ from the indirect improvement produced when $\eta_{pp}$ is more tightly constrained.

A coordinate-robust comparison of the relative information supplied by the joint measurement was obtained from the generalized eigenvalue problem

$$F_{joint} v = \lambda F_{XAS} v.$$

The generalized eigenvalues quantify the relative Fisher-information gain along independent local directions. Because both Fisher matrices undergo the same congruence transformation under a nonsingular local change of coordinates, these generalized eigenvalues are invariant under such reparameterizations. They were therefore used to determine whether

Auger spectroscopy improves all locally reconstructable directions uniformly or predominantly strengthens a specific weakly constrained mode. Spectral derivatives entering the Fisher matrices were evaluated numerically using centered finite differences,

$$\frac{\partial \mu_k}{\partial \theta_\alpha} \approx \frac{\mu_k(\theta + h_\alpha e_\alpha) - \mu_k(\theta - h_\alpha e_\alpha)}{2h_\alpha},$$

about the reference point $(s_\Gamma, \eta_{pp}) = (1,1)$. The baseline perturbation was $h_{s_\Gamma} = h_{\eta_{pp}} = 0.02$. Numerical convergence with respect to the derivative step is assessed separately in Sec. 3.3 using $h = 0.04, 0.02, 0.01, 0.005$.

### 2.4. Robustness Analysis, Reconstruction Test, and Measurement Design

Several additional analyses were performed to distinguish the structural XAS−Auger complementarity from numerical or model-dependent effects. These tests addressed four issues: sensitivity to the detailed Auger spectral representation and finite excitation bandwidth, agreement between local Fisher predictions and explicit noisy-data reconstruction, selected instrumental nuisance parameters, and allocation of a fixed detected-count budget between the two observables.

As a first model-robustness test, an auxiliary kinetic-energy representation of the Auger observable was constructed in which the primary autoionization channels were placed at their calculated electron kinetic energies rather than aggregated solely by ionic final-state energy. This auxiliary representation preserves the same parameterization of the primary decay rates and, in particular, the exact cancellation of the common scale $s_\Gamma$ from the normalized branching fractions. It was used only as a robustness and reconstruction test; the final-state-resolved spectrum defined in Sec. 2.2 remains the baseline Auger model used for the principal Fisher results.

Within this auxiliary representation, the sensitivity to finite excitation bandwidth was examined by introducing an effective Gaussian broadening,

$$\sigma_{eff} = \sqrt{\sigma_A^2 + \sigma_{ph}^2},$$

where the photon contribution was obtained from the specified photon-energy FWHM,

$$\sigma_{ph} = \frac{FWHM_{ph}}{2\sqrt{2In2}}.$$

Photon-bandwidth FWHM values of 0, 0.02, 0.04, 0.06 eV were considered. For each value, the auxiliary Auger spectrum and corresponding Auger-only and joint Fisher matrices were recomputed, while the XAS forward model was kept unchanged. Changes in Fisher rank, CRLBs, and spectral shape were compared with the zero-bandwidth reference. The tested bandwidths are below the excitation-weighted mean intermediate-state width of approximately 0.098 eV and therefore probe a bandwidth range centered on the narrow-band regime considered in Sec. 2.2. This procedure is strictly an effective broadening test within the adopted incoherent line-shape description and should not be interpreted as a coherent finite-bandwidth Kramers−Heisenberg treatment [20].

The local Fisher approximation was tested through explicit Poisson reconstruction using the same auxiliary Auger representation. A total of $N_{MC} = 1000$ independent Poisson realizations were generated at the reference point ($s_\Gamma$, $\eta_{pp}$) = (1,1). Because the normalized Auger observable is independent of $s_\Gamma$, Auger-only reconstructions estimated $\eta_{pp}$, whereas joint XAS−Auger reconstructions estimated both $s_\Gamma$, $\eta_{pp}$. To make the repeated optimizations computationally tractable, second-order local response surfaces were constructed from direct evaluations of the state-resolved forward models. Their accuracy was checked independently against the corresponding direct spectra before use in the Monte Carlo fits. The empirical parameter standard deviations and correlations obtained from the reconstructed ensembles were then compared with the corresponding local Fisher predictions. This Monte Carlo test assesses the local Fisher approximation for the auxiliary representation; it is not used as an independent validation of the final-state-resolved baseline model.

A separate nuisance-parameter analysis examined uncertainty in the calibrated XAS instrumental resolution. The Gaussian XAS width was parameterized as

$$\sigma_{XAS} = q_\sigma \sigma_{XAS}^{(0)},$$

with reference value $q_\sigma = 1$.

Relative calibration uncertainties of 1%, 2%, 5%, and 10% were represented by Gaussian priors on $q_\sigma$. The corresponding prior Fisher contribution,

$$F_{q_\sigma q_\sigma}^{prior} \frac{1}{\sigma_{q_\sigma}^2},$$

was added to the augmented Fisher matrix before marginalization over $q_\sigma$. This test was used to determine how uncertainty in the instrumental Gaussian width propagates into the inferred decay scale and channel-decomposition parameters.

A purely multiplicative gain was also considered as a limited nuisance test. A common gain parameter $g$ was introduced through

$$\mu_\kappa^{XAS} \rightarrow g\,\mu_\kappa^{XAS},$$

and

$$\mu_\iota^{XAS} \rightarrow g\,\mu_\iota^{XAS},$$

so that the parameter vector was augmented from $\theta = (s_\Gamma, \eta_{pp})$ to $\theta' = (s_\Gamma, \eta_{pp}, g)$.

For each fixed-count normalized observable,

$$\sum_\kappa \frac{\partial \mu_\kappa}{\partial \theta_\alpha} = 0$$

for the spectral-shape parameters. Consequently, the Fisher cross terms between a purely multiplicative gain and the normalized shape parameters vanish analytically under the fixed-count construction. The augmented Fisher matrix was nevertheless evaluated numerically as a consistency check. This result applies only to a pure multiplicative gain and does not represent photon-energy calibration error, fluctuations in the incident spectral profile, structured backgrounds, detector nonlinearity, or other correlated experimental systematics.

Finally, the Fisher framework was used to examine the allocation of a fixed total expected detected-count budget,

$$N_{tot} = N_{XAS} + N_{Auger} = 1.8 \times 10^5,$$

equal to the total reference count budget. Under the independent Poisson model, Fisher information scales linearly with expected detected counts. Fisher information per detected count was therefore defined as

$$\tilde{F}_{XAS} = \frac{F_{XAS}^{(0)}}{N_{XAS}^{(0)}}, \quad \tilde{F}_{Auger} = \frac{F_{Auger}^{(0)}}{N_{Auger}^{(0)}}.$$

If $f_A$ denotes the fraction of the detected-count budget assigned to Auger spectroscopy,

$N_{Auger} = f_A N_{tot}$, and $N_{XAS} = (1 - f_A) N_{tot}$, with $0 \leq f_A \leq 1$.

The corresponding Fisher matrix was constructed as

$$F(f_A) = N_{tot}\left[(1 - f_A)\tilde{F}_{XAS} + f_A \tilde{F}_{Auger}\right].$$

The allocation was optimized separately for the marginalized CRLB uncertainties of $s_\Gamma$ and $\eta_{pp}$. In addition, three standard two-parameter design criteria [22] were evaluated in the adopted dimensionless parameterization: D-optimality, obtained by maximizing

$$\det F \;\; ;$$

A-optimality, obtained by minimizing

$$Tr(F^{-1});$$

and E-optimality, obtained by maximizing

$$\lambda_{min}(F).$$

These criteria quantify different aspects of the local uncertainty geometry and therefore need not select the same allocation. The resulting optimization concerns the distribution of a fixed number of expected detected events between the two observables; it should not be interpreted directly as an optimization of beam time, incident photon flux, or experimental cost.

Together, these analyses probe model robustness, reconstruction consistency, nuisance-parameter sensitivity, and detected-count allocation. Their purpose is not to establish a complete experimental error model, but to test whether the central degeneracy-breaking mechanism identified by the Fisher analysis persists under selected perturbations of the adopted measurement model and to determine how the complementary information can be distributed under a fixed detected-count constraint.

# 3. Results and Discussion

## 3.1. State-Resolved $Ar^+$ Spectral Structure and Final-State Lifetime Broadening

The state-resolved FAC calculation yields a dense $2p \rightarrow 3d$ excitation manifold. The atomic-structure calculation contains 65 core-excited fine-structure levels distributed between 251.679 and 260.497 eV on the FAC energy scale. From the two fine-structure components of the $2p^6 3s^2 3p^5$ ground configuration, 76 electric-dipole (E1) transitions reach 47 distinct core-excited levels; the remaining 18 calculated levels carry zero E1 excitation weight within the adopted model. The resulting absorption spectrum (Fig. 1) is therefore formed by a superposition of partially overlapping resonances rather than by an isolated lifetime-broadened feature. Each contributing resonance carries its own excitation strength and state-specific decay width, so the observable XAS profile reflects simultaneously the distribution of oscillator strength, the density of intermediate states, natural lifetime broadening, and instrumental resolution.

The decay manifold is considerably richer (Fig. 2). Primary autoionization from the $2p^5 3s^2 3p^5 3d$ states into the five included $Ar^{2+}$ configuration groups yields 7397 resolved decay channels. The decay is strongly spectator dominated: spectator channels account for approximately 95.49% of the excitation-weighted primary Auger strength, with the spectator-$pp$ group alone contributing 63.64%, followed by spectator-$sp$ at 28.12% and spectator-$ss$ at 3.73%. The two participator groups contribute only 1.97% and 2.54%, respectively. This pronounced channel hierarchy is important for the subsequent inference problem because the parameter $\eta_{pp}$ selectively perturbs the dominant spectator-$pp$ contribution rather than a weak peripheral pathway.

The core-excited states are overwhelmingly autoionizing within the radiative channel set included in the present calculation. The excitation-weighted mean Auger-only natural FWHM is $\langle \Gamma_{Auger} \rangle_{XAS} = 0.098268$ eV, whereas inclusion of the calculated E1 radiative contribution increases the corresponding excitation-weighted mean total natural width only to $\langle \Gamma_{tot} \rangle_{XAS} = 0.098296$ eV. The difference is $2.81 \times 10^{-5}$ eV. Consistently, the XAS-weighted radiative fraction associated with the explicitly included E1 channels is only $3.38 \times 10^{-4}$, compared with an Auger fraction of 0.999662. Thus, within the present atomic model, the natural widths of the $2p \rightarrow 3d$ intermediate states are controlled almost entirely by autoionization, while the included radiative contribution has a negligible effect on the XAS lifetime broadening.

A different situation arises for the ionic states populated by primary Auger decay. Of the 115 $Ar^{2+}$ fine-structure levels generated by the five final-state configuration groups, 46 possess nonzero secondary-autoionization rates into the included $Ar^{3+}$ manifold. Their calculated nonzero final-state natural FWHM values span $1.65 \times 10^{-5}$ eV $\leq \Gamma_f \leq 0.897$ eV. Primary transitions terminating in these autoionizing $Ar^{2+}$ levels account for approximately 23.17% of the excitation-weighted primary decay strength. Within this unstable subset, the spectral-weighted mean final-state width is approximately $\langle \Gamma_f \rangle_{unstable} \approx 0.409$ eV. Final-state lifetime broadening is therefore highly nonuniform across the decay manifold and is concentrated in a substantial spectroscopically weighted subset of the final states.

As an external consistency check on the calculated decay branching, primary transitions terminating in the 46 autoionizing $Ar^{2+}$ levels were classified as feeding subsequent $Ar^{3+}$ production, while the remaining primary transitions terminate in $Ar^{2+}$ levels that are stable against secondary autoionization within the included configuration space. Under this simplified sequential-decay classification, the excitation-weighted fractions over the complete calculated excitation manifold are 76.8% for $Ar^{2+}$ and 23.2% for $Ar^{3+}$. Restricting the calculation to the 252.5−256.5 eV interval on the FAC transition-energy scale, which contains approximately 91.9% of the total excitation weight, gives 77.6% $Ar^{2+}$ and 22.4% $Ar^{3+}$.

Experimental measurements in the corresponding nominal $2p \rightarrow 3d$ energy region report closely comparable product-charge fractions: 77.0% $Ar^{2+}$, 21.8% $Ar^{3+}$, and approximately 1.2% $Ar^{4+}$ in the high-resolution study of Müller et al. [17], while the earlier measurements of Blancard et al. [16] give approximately 75%, 24%, and 1% for the corresponding single-, double-, and triple- ionization products. The present calculation does not explicitly resolve the complete higher-charge-state cascade, including the $Ar^{4+}$ and higher product channels observed experimentally. Furthermore, the calculated and experimental energy scales are not independently aligned here. This comparison is therefore intended only as an aggregate consistency check on the calculated partition of primary decay strength between stable and autoionizing $Ar^{2+}$ final states; it does not constitute a complete prediction of experimental charge-state yields or a level-by-level validation of the calculated final-state widths.

Within the effective final-state-resolved Auger model defined in Sec. 2.2, a common multiplication of the primary Auger rates by $s_\Gamma$ leaves the normalized branching fractions unchanged, while the Lorentzian widths assigned to the Auger features are fixed by the calculated $Ar^{2+}$ final-state lifetimes. Under the fixed-excitation-weight and unit non-radiative-yield assumptions of the adopted model, the normalized Auger observable therefore carries no direct Fisher information on $s_\Gamma$. By contrast, the channel-selective parameter $\eta_{pp}$ changes the relative branching into the dominant spectator-$pp$ channels and consequently modifies the normalized Auger spectral shape. The two observables therefore probe complementary aspects of the intermediate-state decay scale and its channel-weighted combinations, whereas the effective final-state-resolved Auger observable supplies direct information on the channel decomposition. This information complementarity is quantified in Sec. 3.2.

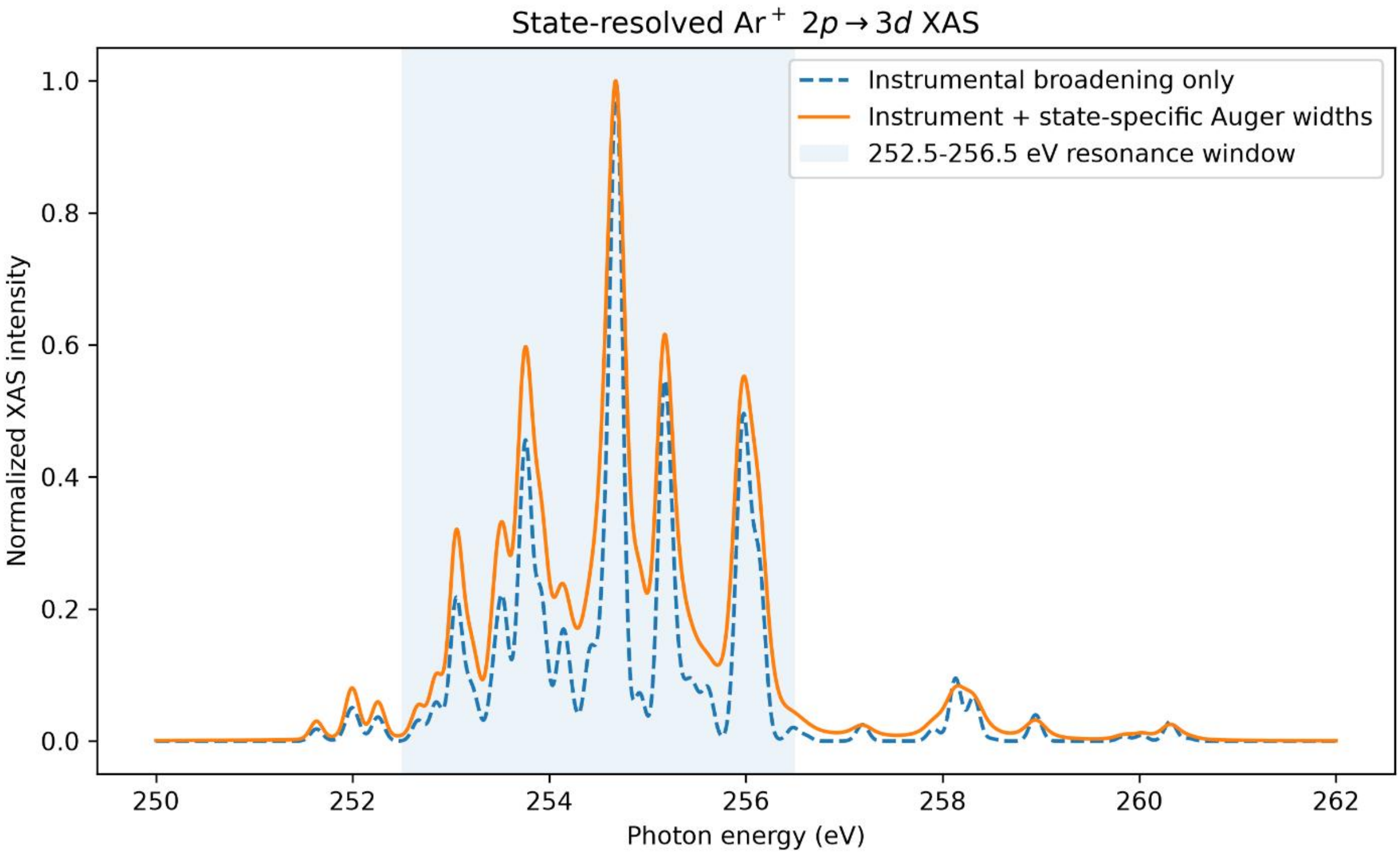

Figure 1: Calculated state-resolved $Ar^+$ 2p→3d X-ray absorption spectrum obtained from FAC transition energies and oscillator strengths. The dashed curve represents Gaussian instrumental broadening only (FWHM = 0.140 eV), whereas the solid curve includes the state-specific primary Auger lifetime widths through Voigt profiles. Both spectra are normalized to their maximum intensity. The shaded region indicates the 252.5–256.5 eV resonance window used for the restricted excitation-weight analysis.

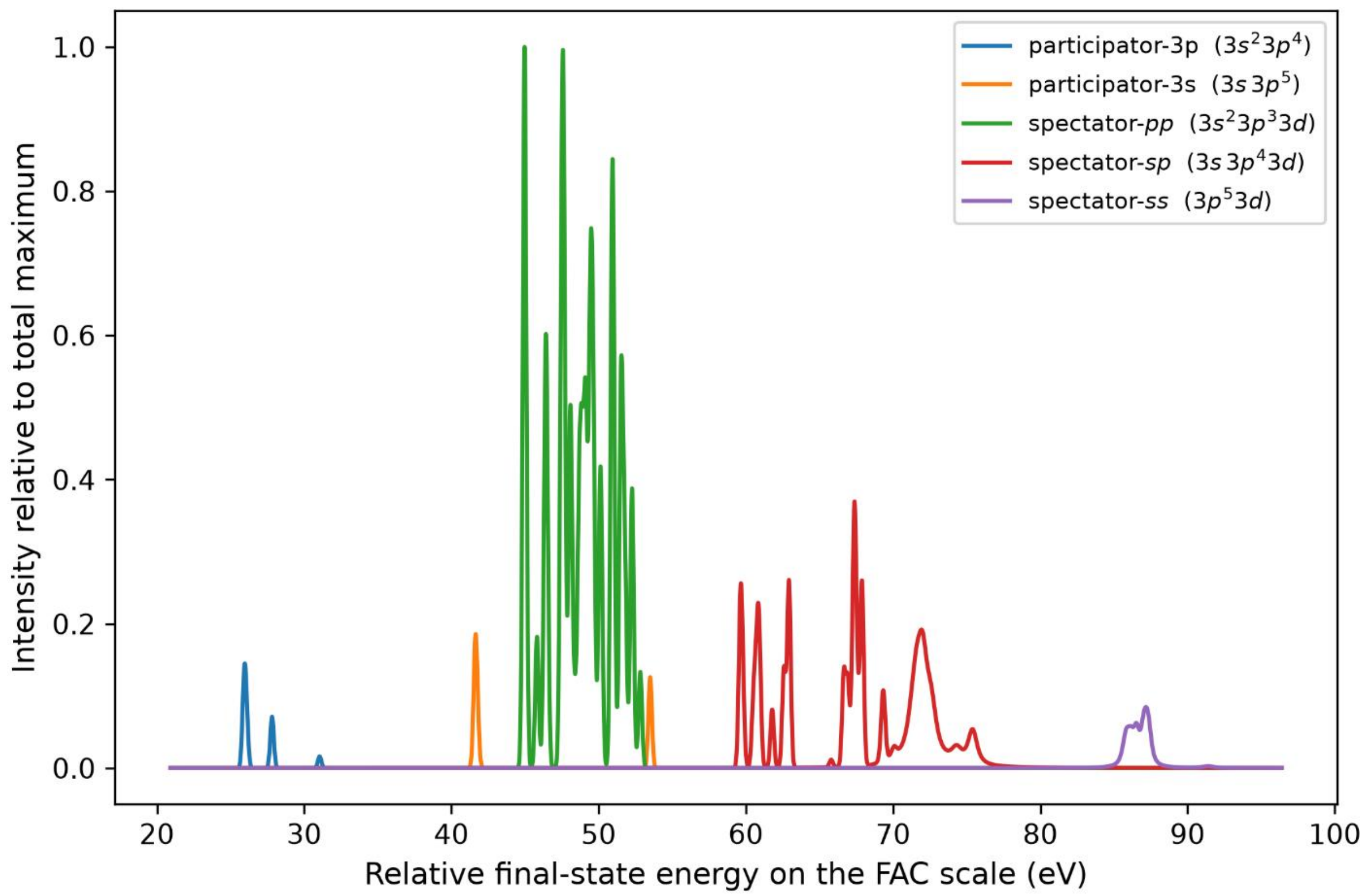

Figure 2: Calculated effective final-state-resolved resonant-Auger spectrum of $Ar^+$, decomposed into five primary autoionization channel groups: participator-3p, participator-3s, spectator-pp, spectator-sp, and spectator-ss. The spectrum combines excitation-weighted FAC branching fractions into 115 $Ar^{2+}$ final levels. Each contribution includes Gaussian analyzer broadening (σ = 0.12 eV) and the calculated final-state lifetime broadening. Intensities are normalized to the maximum of the total spectrum. The horizontal axis represents relative ionic final-state energy on the FAC scale, rather than directly measured electron kinetic or binding energy.

## 3.2. Parameter-Specific Information, Physical Reparameterization, and Degeneracy Breaking

The final-state-resolved model reveals a sharp distinction between sensitivity to the overall decay scale and sensitivity to its channel decomposition. In the adopted parameterization, $\theta = (s_\Gamma, \eta_{pp})$, both parameters modify the state-dependent intermediate-state widths entering XAS. Their spectral derivatives, however, are nearly collinear, so the two effects are difficult to separate despite substantial overall XAS sensitivity.

At the reference count budget, the XAS Fisher matrix is

$$F_{XAS} = \begin{pmatrix} 7053.478 & 4542.016 \\ 4542.016 & 2925.989 \end{pmatrix}.$$

Its eigenvalues are $\lambda_{min}^{XAS} = 0.8508$, $\lambda_{max}^{XAS} = 9978.62$, giving $\kappa_{XAS} \approx 1.17 \times 10^4$, $\rho_{XAS} = -0.999794$. XAS is therefore formally full rank but almost one-dimensional in the two-parameter space. The corresponding marginalized CRLB standard uncertainties are $\sigma_{XAS}(s_\Gamma) = 0.5871$, $\sigma_{XAS}(\eta_{pp}) = 0.9115$. These large marginalized uncertainties do not indicate weak XAS sensitivity. Rather, they arise because most of the information carried by the two spectral derivatives lies along nearly the same local parameter direction, so that separating the two contributions incurs a severe correlation penalty.

The physical origin of this near-degeneracy follows directly from the state-resolved decay rates. Define the spectator-$pp$ fraction of intermediate state $i$ as

$$p_i = \frac{A_{i,pp}^{(0)}}{A_i^{Auger,(0)}}.$$

At the reference point $(s_\Gamma, \eta_{pp}) = (1, 1)$,

$$\frac{\partial \Gamma_i}{\partial s_\Gamma} = \hbar A_i^{Auger,(0)},$$

whereas

$$\frac{\partial \Gamma_i}{\partial \eta_{pp}} \hbar A_{i,pp}^{(0)} = p_i \hbar A_i^{Auger,(0)}.$$

If $p_i$ were identical for all populated intermediate states, the two linewidth derivatives would be exactly proportional. XAS could then constrain only one local combination of $s_\Gamma$ and $\eta_{pp}$, and its two-parameter Fisher matrix would be rank one. In the calculated manifold, the excitation-weighted mean spectator-$pp$ fraction is approximately 0.63636, while its weighted state-to-state standard deviation is only 0.0276, corresponding to about 4.3% of the mean. This limited state-to-state variation breaks the proportionality only weakly and accounts for the small but finite minimum XAS eigenvalue. The near-singularity is therefore a consequence of the similar decay-channel composition of the populated core-excited states rather than of weak overall spectral sensitivity.

The normalized final-state-resolved Auger observable has the complementary information structure. Because the common scale $s_\Gamma$ cancels exactly from the normalized primary branching fractions, while the final-state linewidths are fixed independently by the calculated $Ar^{2+}$ lifetimes, under the assumptions of the adopted forward model,

$$\frac{\partial S_{Auger}}{\partial s_{\Gamma}} = 0.$$

By contrast, $\eta_{pp}$ selectively modifies the spectator-$pp$ branching and consequently redistributes normalized spectral weight among the ionic final states,

$$\frac{\partial S_{Aguer}}{\partial \eta_{pp}} \neq 0.$$

The corresponding final-state-resolved Auger Fisher matrix is therefore rank one,

$$F_{Auger} = \begin{pmatrix} 0 & 0 \\ 0 & 18364.548 \end{pmatrix},$$

yielding $\sigma_{Auger}(\eta_{pp}) = 7.379 \times 10^{-3}$.

No finite Auger-only two-parameter CRLB exists for $s_{\Gamma}$, because the normalized Auger observable contains no direct information along that parameter coordinate.

As an analytic upper-bound check, consider an idealized fully labeled measurement in which the intermediate-state and primary-channel identities are retained. At the reference point,

$$F_{\eta\eta}^{full-label} = N_{Auger} \sum_{i} w_i^{exc} p_i (1 - p_i) = 18451.73.$$

The adopted final-state-resolved spectrum provides $F_{\eta\eta}^{Auger} = 18364.55$, and therefore retains

$$\frac{F_{\eta\eta}^{Auger}}{F_{\eta\eta}^{Auger-label}} = 0.99528,$$

or approximately 99.53%, of this idealized labeled-channel information. Relative to the looser Jensen bound,

$$N_{Auger} \bar{B}_{pp} (1 - \bar{B}_{pp}) = 18512.50,$$

the retained fraction is approximately 99.20%. Thus, the strong Auger sensitivity to $\eta_{pp}$ survives aggregation into the final-state-resolved spectrum and is not an artifact of treating the 7397 primary autoionization channels as independently observed quantities.

When the XAS and Auger information are combined,

$$F_{joint} = F_{XAS} + F_{Auger} = \begin{pmatrix} 7053.478 & 4542.016 \\ 4542.016 & 21290.537 \end{pmatrix}.$$

Auger changes only the $\eta_{pp}\eta_{pp}$ element; $F_{ss}$ and $F_{s\eta}$ remain identical to their XAS-only values. Nevertheless, the joint inverse problem becomes well conditioned, $\lambda_{min}^{joint} = 5727.88$, $\lambda_{max}^{joint} = 22616.14$, with $\kappa_{joint} = 3.95$, $\rho_{joint} = -0.371$. The local information geometry is summarized in Table 1.

Table 1: Local Fisher-information structure at the reference count budget, $N_{XAS} = 10^5$ and $N_{Auger} = 8 \times 10^4$.

| Quantity | XAS | Auger | XAS + Auger |
|---|---|---|---|
| $F_{ss}$ | 7053.478 | 0 | 7053.478 |
| $F_{s\eta}$ | 4542.016 | 0 | 4542.016 |
| $F_{\eta\eta}$ | 2925.989 | 18364.548 | 21290.537 |
| $\lambda_{min}$ | 0.8508 | 0 | 5727.88 |
| $\lambda_{max}$ | 9978.62 | 18364.548 | 22616.14 |
| $\kappa$ | $1.17 \times 10^4$ | −(rank one) | 3.95 |
| $\rho$ | −0.999794 | − | −0.371 |
| $F_{eff}(s_\Gamma)$ | 2.901 | 0 | 6084.51 |

The mechanism by which Auger improves inference of the common primary-rate scale, despite carrying no direct information on $s_\Gamma$, is exposed by the Schur complement. After marginalizing over $\eta_{pp}$,

$$F_{s_\Gamma}^{eff} = F_{ss} - \frac{F_{s\eta}^2}{F_{\eta\eta}}.$$

For XAS alone, $F_{s_\Gamma XAS}^{eff} = 2.901$, because the correlation penalty almost completely cancels the direct $F_{ss}$ contribution. After adding Auger, $F_{s_\Gamma,joint}^{eff} = 6084.51$. The effective Fisher information on $s_\Gamma$ therefore increases by a factor of approximately $2.10 \times 10^3$, so that the marginalized standard uncertainty of $s_\Gamma$ becomes smaller by a factor of 45.79. Auger thus improves inference of $s_\Gamma$ indirectly: by tightly constraining $\eta_{pp}$, it removes the correlated nuisance degree of freedom that makes the common decay-rate scale poorly identifiable from XAS alone.

The physical significance of this degeneracy breaking becomes clearer after propagating the CRLB covariance-bound matrix to quantities associated with the total decay width and its decomposition into partial-rate contributions. The corresponding local CRLB standard uncertainties are summarized in Table 2, and their fractional reductions are shown in Fig. 3.

Table 2: Local CRLB standard uncertainties of the model parameters and derived physical quantities.

| Quantity | Reference | XAS | XAS + Auger | Reduction |
|---|---|---|---|---|
| $s_\Gamma = s_{non-pp}$ | 1 | 0.5871 | 0.01282 | 97.82% |
| $\eta_{pp}$ | 1 | 0.9115 | 0.007379 | 99.19% |
| $s_{pp} = s_\Gamma \eta_{pp}$ | 1 | 0.3248 | 0.01219 | 96.25% |
| $\bar{\Gamma}_{tot}$(eV) | 0.098296 | $1.178 \times 10^{-3}$ | $1.170 \times 10^{-3}$ | 0.70% |
| $\bar{B}_{pp}$ | 0.636359 | 0.2102 | $1.702 \times 10^{-3}$ | 99.19% |

The excitation-weighted total natural width is therefore already constrained by XAS to a local fractional CRLB of approximately 1.20% under the adopted count and resolution model. Adding Auger changes this only slightly, reducing its standard uncertainty by 0.70%. By contrast, the spectator-$pp$ branching fraction and the two partial-rate scales gain dramatically, with standard-uncertainty reductions of approximately 96–99%. The complementarity is therefore associated primarily with resolving the channel decomposition rather than with refining the total natural width. A coordinate-robust summary follows from the generalized eigenvalue problem

$$F_{joint} v = \lambda F_{XAS} v.$$

Because $F_{joint} - F_{XAS} = F_{Auger}$ is a rank update, one generalized eigenvalue is exactly $\lambda_1 = 1$. The second is $\lambda_2 = 1 + F_{\eta\eta}^{Auger} (F_{XAS}^{-1})_{\eta\eta} \approx 1.5259 \times 10^4$.

Thus, one generalized-information mode is left unchanged, while the complementary mode is enhanced by more than four orders of magnitude in Fisher information. On the corresponding standard-deviation scale, the enhancement is $\sqrt{\lambda_2} \approx 123.5$. It is important to distinguish this generalized eigenmode from the linear parameter combination whose propagated CRLB variance is unchanged by the Auger update. In the adopted $(s_\Gamma, \eta_{pp})$ coordinates, the latter is proportional to

$$s_\Gamma + \frac{F_{s_\eta}^{XAS}}{F_{ss}^{XAS}} \eta_{pp} \approx s_\Gamma + 0.644\eta_{pp}.$$

The corresponding spectator-$pp$ coefficient in the local variation of the excitation-weighted total width is approximately 0.6455. The near coincidence of these two combinations explains why the total-width uncertainty changes by only 0.70%, even though the complementary generalized-information mode is enhanced extremely strongly.

Locally, any differentiable derived quantity that depends only on $\eta_{pp}$ and has nonzero derivative inherits the same relative CRLB improvement as $\eta_{pp}$, because its Jacobian factor cancels in the ratio of XAS-only to joint standard uncertainties. Consequently,

$$\frac{\sigma_{\eta_{pp}}^{XAS}}{\sigma_{\eta_{pp}}^{joint}} = \sqrt{\lambda_2},$$

and the corresponding fractional reduction is

$$1 - \frac{1}{\sqrt{\lambda_2}} \approx 99.19\%.$$

This explains why $\eta_{pp}$ and the excitation-weighted branching fraction $\bar{B}_{pp}$ exhibit the same 99.19% reduction in Table 2.

The two large improvement factors appearing in the analysis quantify different aspects of the same degeneracy-breaking mechanism. The factor 45.79 refers to the reduction of the marginalized standard uncertainty of the specific coordinate $s_\Gamma$ after $\eta_{pp}$ is constrained, whereas the factor 123.5 characterizes the maximally enhanced generalized-information mode. There is therefore no requirement for the two factors to coincide.

Overall, XAS already constrains the excitation-weighted total lifetime broadening at approximately the percent level, but it cannot reliably decompose that width into its underlying decay-channel contributions under the adopted measurement model. The normalized final-state-resolved Auger observable supplies strongly complementary channel-selective information. Its principal role is therefore not to refine an already well-constrained total width, but to convert a nearly degenerate two-parameter inverse problem into a well-conditioned reconstruction of the underlying partial decay scales.

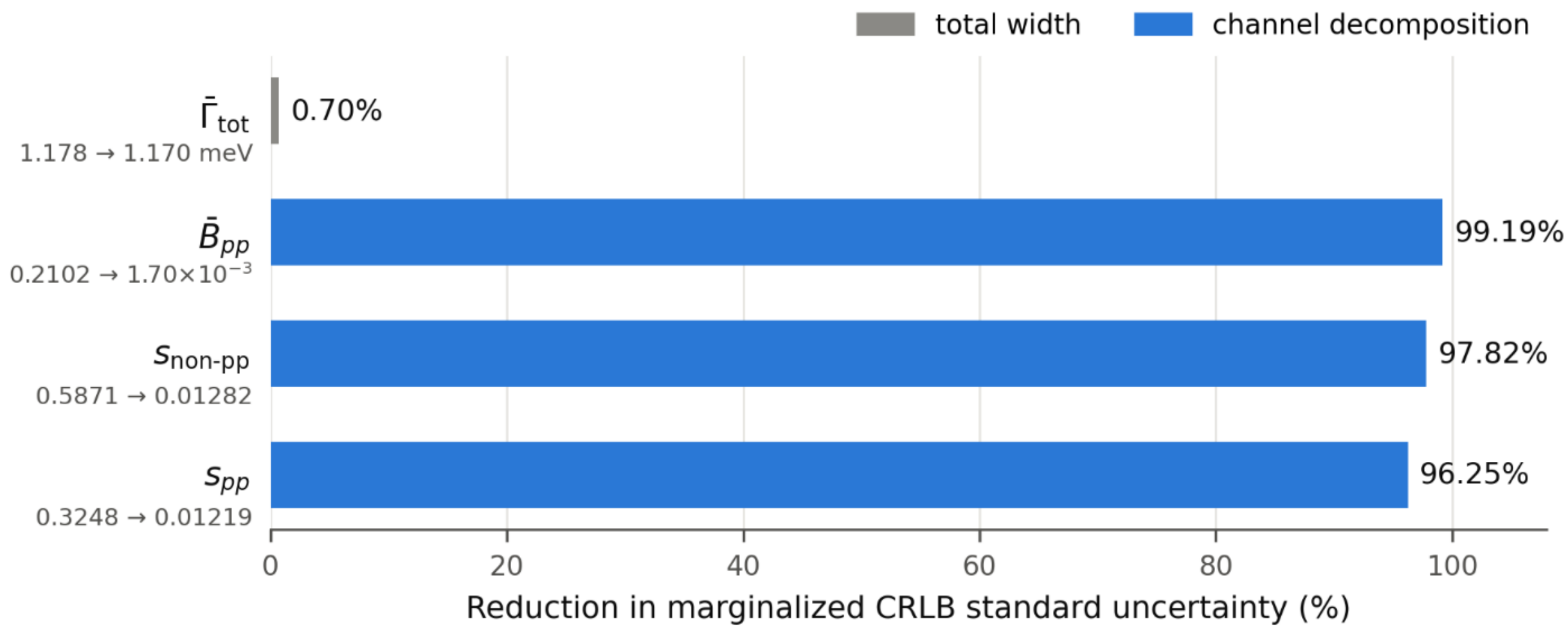

Figure 3: *Reduction in the marginalized Cramér–Rao lower-bound (CRLB) standard uncertainties when final-state-resolved Auger spectroscopy is added to XAS, relative to XAS alone. The analysis uses expected detected-count budgets of* $N_{XAS} = 10^5$ *and* $N_{Auger} = 8 \times 10^4$. *The excitation-weighted total natural width (gray) shows only a 0.70% reduction, whereas the spectator-pp branching fraction and the non-pp and pp partial-rate scale parameters (blue) exhibit reductions of 99.19%, 97.82%, and 96.25%, respectively. Values beside the bars indicate the XAS-only →joint CRLB standard uncertainties, expressed in meV for the total width and in dimensionless units for the other quantities. The contrast demonstrates that the principal information gain from Auger spectroscopy concerns decay-channel decomposition rather than refinement of the total natural width.*

## 3.3. Numerical Validation and Robustness of the Information Geometry

Because the local identifiability analysis relies on numerical derivatives of state-resolved spectral models, the stability of the Fisher information with respect to the finite-difference step was tested explicitly. Symmetric perturbations $h = 0.04, 0.02, 0.01, 0.005$ were evaluated for both the XAS and final-state-resolved Auger calculations. For the normalized final-state-resolved Auger observable, the common primary-rate scale $s_\Gamma$ cancels exactly from the branching fractions. Consequently, $F^{Auger}_{s_\Gamma s_\Gamma} = F^{Auger}_{s_\Gamma \eta_{pp}} = 0$ analytically. The rank-one structure of the Auger Fisher matrix is therefore a structural consequence of the forward model rather than a numerical convergence result. The finite-difference test instead probes the stability of the nonzero element $F^{Auger}_{\eta_{pp}\eta_{pp}}$ and of the resulting joint CRLBs. Relative to the $h = 0.02$ baseline, $F^{Auger}_{\eta_{pp}\eta_{pp}}$ varied by less than 0.10% across the tested range, while the joint marginalized standard uncertainties varied by less than approximately 0.015% for $s_\Gamma$ and 0.049% for $\eta_{pp}$. The numerical Fisher values are therefore stable with respect to the finite-difference step over the tested interval.

A separate robustness test examined whether the central information-geometric result depends strongly on the detailed representation of the Auger spectrum. An auxiliary kinetic-energy representation was constructed in which the primary autoionization channels were placed at their calculated electron kinetic energies rather than being aggregated through the final-state-energy representation used in the baseline analysis. This alternative model preserves the same rate parameterization and, in particular, the exact cancellation of $s_\Gamma$ from the normalized Auger branching fractions. It yields $F^{Auger}_{\eta_{pp}\eta_{pp}} = 17625.4$, compared with 18364.5 for the final-state-resolved baseline. The Auger information therefore differs by approximately 4.2% between the two representations. The corresponding joint marginalized uncertainties remain close, $\sigma(s_\Gamma) = 0.01286$ vs. 0.01282, and $\sigma(\eta_{pp}) = 0.007532$ vs. 0.007379, for the auxiliary and final-state-resolved representations, respectively. Thus, the detailed spectral representation modifies the quantitative information content only modestly and does not alter the underlying degeneracy-breaking mechanism.

The local Fisher prediction was further tested within the auxiliary representation by explicit noisy-data reconstruction. A total of $N_{MC} = 10^3$ independent Poisson realizations were generated at the reference point $(s_\Gamma, \eta_{pp}) = (1, 1)$. Because the normalized Auger spectrum is independent of $s_\Gamma$, the Auger-only reconstructions estimated $\eta_{pp}$, whereas the joint XAS−Auger reconstruction estimated both parameters. All optimization runs converged. For the Auger-only reconstruction,

$$\frac{\sigma_{MC}(\eta_{pp})}{\sigma_{CRLB}(\eta_{pp})} = 0.998.$$

For the joint reconstruction, the corresponding ratios were 0.975 for $s_\Gamma$ and 0.997 for $\eta_{pp}$. The empirical parameter correlation, $\rho_{MC} = -0.397$, was close to the local Fisher prediction for the same auxiliary representation, $\rho_{Fisher} = -0.377$. for $N_{MC} = 1000$, the relative sampling uncertainty of an empirical standard deviation is approximately 2.2%, while the sampling uncertainty of the correlation coefficient is of order 0.03. The observed deviations are therefore compatible with the statistical resolution of the Monte Carlo ensemble. This test supports the local Fisher approximation for the auxiliary spectral representation; it does not constitute an independent validation of the final-state-resolved baseline line-shape model.

To make the $10^3$-realization reconstruction computationally tractable, the repeated optimizations used second-order local response surfaces constructed from direct evaluations of the state-resolved forward models. The maximum total-variation distances between the surrogate and direct spectra were $4.63 \times 10^{-6}$ for XAS and $2.21 \times 10^{-6}$ for Auger. These deviations are negligible relative to the spectral variations and Poisson fluctuations sampled in the reconstruction, indicating that surrogate approximation error does not control the Monte Carlo results.

Sensitivity to finite excitation bandwidth was examined separately within the same auxiliary effective line-shape framework. Gaussian photon-bandwidth FWHM values of 0, 0.02, 0.04, 0.06 eV were incorporated through the effective Gaussian broadening defined in Sec. 2.4. Across this tested range, the Auger observable retained its exact structural independence from $s_\Gamma$, while the joint Fisher matrix remained full rank. The maximum degradation of the CRLB standard uncertainty was approximately 0.0094% for the Auger determination of $\eta_{pp}$ and 0.0013% for the joint determination of $s_\Gamma$. The largest total-variation distance from the zero-bandwidth Auger spectrum was $2.80 \times 10^{-3}$. Within the adopted incoherent effective-broadening model, the local information geometry is therefore only weakly affected by the tested additional broadening. This result should not be interpreted as a coherent finite-bandwidth Kramers–Heisenberg calculation, which would require excitation–decay amplitudes and their relative phases beyond the rate-based model used here [4,20].

A separate nuisance-parameter analysis examined uncertainty in the calibrated XAS instrumental resolution. The Gaussian XAS width was scaled by the parameter $q_\sigma$ introduced in Sec. 2.4, with relative calibration priors of 1%, 2%, 5%, and 10%. Using the final-state-resolved Auger Fisher contribution, marginalization over $q_\sigma$ increased the joint CRLB standard uncertainty of $s_\Gamma$, relative to the fixed-resolution result, by factors of 1.108, 1.206, 1.277, 1.291, respectively. By contrast, the corresponding uncertainty of $\eta_{pp}$ changed by less than 0.001% over the entire tested range. This asymmetry follows directly from the information geometry: the XAS instrumental width is partially confounded with the intermediate-state decay-width scale, whereas the strong Auger constraint on the channel-selective parameter does not depend on the XAS resolution parameter in the adopted model. Resolution uncertainty therefore degrades primarily the inference of the absolute decay scale while leaving the Auger-driven determination of $\eta_{pp}$ essentially unchanged.

A purely multiplicative gain was considered as a separate, limited nuisance test. Under the normalized spectral-shape construction with parameter-independent baseline expected count totals,

$$\sum_{\kappa} \frac{\partial \mu_\kappa}{\partial \theta_\alpha} = 0$$

for the physical shape parameters. The Fisher cross terms between a common multiplicative gain and these normalized shape parameters therefore vanish analytically. The corresponding numerical cross terms were of order $10^{-10}$ in unnormalized Fisher entries, while their normalized coupling coefficients were of order $10^{-15}$. Marginalization over the gain left the physical CRLBs unchanged to numerical precision. This orthogonality is a consequence of the adopted spectral-shape normalization and should not be interpreted as a general statement of instrumental robustness. In particular, photon-energy calibration errors, fluctuations in the incident spectral profile, structured backgrounds, detector nonlinearity, energy-dependent detection efficiency, and absolute-yield systematics are not represented by this gain parameter and would require separate nuisance models.

Taken together, these tests distinguish three logically different levels of validation. First, the finite-difference study establishes the numerical stability of the Fisher values and joint CRLBs with respect to the derivative step; it does not establish the rank-one structure of the Auger information, which follows analytically from the normalized branching model. Second, the alternative Auger representation and its Monte Carlo reconstruction show that the central degeneracy-breaking mechanism survives a modest change in the spectral forward model and remains consistent with explicit noisy-data reconstruction in the local regime. Third, the finite-bandwidth and XAS-resolution studies test selected perturbations of the measurement model rather than a complete experimental uncertainty budget. Within these tested conditions, the central result of Sec. 3.2 remains unchanged: the normalized Auger observable contains no direct information on the common primary-rate scale $s_\Gamma$, but its strong constraint on $\eta_{pp}$ removes the near-degenerate channel-decomposition degree of freedom left by XAS.

## 3.4. Information-Optimal XAS−Auger Count Allocation

The preceding analysis shows that XAS and final-state-resolved Auger spectroscopy constrain different directions of the same local parameter space. This complementarity has a direct consequence for measurement design: under a fixed expected detected-count budget, the allocation of counts between the two observables need not be optimal for all inferred quantities simultaneously. The total expected detected-count budget was fixed at $N_{tot} = N_{XAS} +$

$N_{Auger} = 1.8 \times 10^5$, equal to that of the reference calculation. The Auger fraction was defined as

$$f_A = \frac{N_{Auger}}{N_{tot}},$$

with $N_{Auger} = f_A N_{tot}$, $N_{XAS} = (1 - f_A) N_{tot}$. Under the independent Poisson model adopted here, Fisher information scales linearly with the expected detected counts. The analysis therefore compares alternative allocations of a fixed detected-count budget; it should not be interpreted as an optimization of beam time, incident photon flux, or experimental cost.

The optimal allocation depends strongly on the quantity targeted by the inference. For the marginalized uncertainty in the global primary-rate scale $s_\Gamma$, the minimum occurs at $f_A = 0.2630$, corresponding to approximately $N_{Auger} = 4.73 \times 10^4$, $N_{XAS} = 1.33 \times 10^5$. At this allocation, $\sigma(s_\Gamma) = 0.01204$, compared with 0.01282 for the reference allocation $N_{XAS} = 10^5$, $N_{Auger} = 8 \times 10^4$, for which $f_A = 0.4444$. Reallocating the same total expected detected-count budget therefore reduced the marginalized CRLB standard uncertainty in $s_\Gamma$ by approximately 6.1%.

The asymmetric optimum follows directly from the information geometry established in Sec. 3.2. XAS provides the direct sensitivity to $s_\Gamma$, whereas the normalized Auger observable carries no direct Fisher information on this common rate scale. A finite Auger contribution nevertheless constrains $\eta_{pp}$, thereby reducing the correlation penalty that makes the marginalized determination of $s_\Gamma$ poorly conditioned when both parameters are allowed to vary. Once the channel-selective direction is sufficiently constrained, additional Auger counts provide diminishing indirect benefit to $s_\Gamma$, whereas additional XAS count continue to strengthen the direction containing direct rate-scale information.

This behavior is particularly clear in the XAS-only limit. Assigning the entire $1.8 \times 10^5$-count budget to XAS increases the magnitude of the Fisher information but leaves its local geometry unchanged. The condition number and correlation remain approximately $\kappa_{XAS} = 1.17 \times 10^4$, $\rho_{XAS} = -0.9998$, while the marginalized uncertainties become $\sigma(s_\Gamma) = 0.438$, $\sigma(\eta_{pp}) = 0.679$. Thus, increasing the XAS count budget alone scales the information already present without supplying an independent channel-selective constraint. The limitation is therefore geometric rather than simply a consequence of insufficient counting statistics.

A different boundary solution is obtained when $\eta_{pp}$ alone is treated as the target parameter. Because the normalized Auger observable is directly sensitive to the relative spectator-$pp$ branching structure, the single-parameter optimum occurs at $f_A = 1$, for which $\sigma(\eta_{pp}) = 4.92 \times 10^{-3}$. An Auger-only design, however, cannot simultaneously determine $s_\Gamma$, because the normalized Auger spectrum is invariant under a common rescaling of all primary Auger rates. Optimization for one identifiable quantity can therefore lead to a design that is singular for the full two-parameter problem.

For simultaneous estimation of both parameters, standard multiparameter design criteria were evaluated in the adopted dimensionless $(s_\Gamma, \eta_{pp})$ parameterization. D-optimality maximizes $\det F$, equivalently minimizing $\det(F^{-1})$ and the area of the local covariance ellipse at a fixed confidence level in this two-parameter representation. It gives $f_A = 0.5000$, with $\sigma(s_\Gamma) = 0.01333$, $\sigma(\eta_{pp}) = 0.00696$. The Fisher determinant is approximately 1.25% larger than at the reference allocation. A-optimality minimizes the sum of marginalized variances, $\mathrm{Tr}(F^{-1})$, and gives $f_A = 0.3973$, with $\sigma(s_\Gamma) = 0.01249$, $\sigma(\eta_{pp}) = 0.00780$, E-optimality instead maximizes the weakest Fisher-information direction, $\lambda_{min}(F)$, and yields $f_A = 0.3684$, for which $\lambda_{min} = 5.91 \times 10^3$, with $\sigma(s_\Gamma) = 0.01233$, $\sigma(\eta_{pp}) = 0.00810$. The reference allocation, $f_A = 0.4444$, lies between these multiparameter optima. The principal count-allocation results are summarized in Table 3.

Table 3: Optimal allocations of a fixed budget of 1.8 ×10[5] expected detected counts between XAS and Auger, together with the corresponding marginalized CRLB standard uncertainties.

| Design objective | $f_A$ | $N_{Auger}$ | $N_{XAS}$ | $\sigma(s_\Gamma)$ | $\sigma(\eta_{pp})$ |
|---|---|---|---|---|---|
| $s_\Gamma$-specific | 0.2630 | $4.73\times10^4$ | $1.33\times10^5$ | 0.01204 | 0.00959 |
| E-optimal | 0.3684 | $6.63\times10^4$ | $1.14\times10^5$ | 0.01233 | 0.00810 |
| A-optimal | 0.3973 | $7.15\times10^4$ | $1.08\times10^5$ | 0.01249 | 0.00780 |
| Reference | 0.4444 | $8.00\times10^4$ | $1.00\times10^5$ | 0.01282 | 0.00738 |
| D-optimal | 0.5000 | $9.00\times10^4$ | $9.00\times10^4$ | 0.01333 | 0.00696 |

The absence of a unique optimal allocation is not a numerical ambiguity but a consequence of the anisotropic information geometry. Parameter-specific optimization minimizes the uncertainty of a selected quantity; D-optimality minimizes the generalized covariance volume; A-optimality minimizes the sum of marginal variances; and E-optimality protects the weakest constrained local direction. Because these criteria answer different design questions, they need not select the same count allocation.

More importantly, the optimization distinguishes information necessity from count dominance. In the two-parameter problem, Auger information is particularly effective at resolving the near-degenerate channel-decomposition direction left poorly constrained by XAS, even though the Auger observable carries no direct Fisher information on $s_\Gamma$ and the $s_\Gamma$-specific optimum assigns only about 26% of the detected counts to Auger. Conversely, XAS provides the direct intermediate-state width sensitivity but cannot efficiently separate the underlying decay components without the channel-selective constraint supplied by Auger.

The XAS−Auger problem is therefore not a competition between two measurements for the larger count share. It is an allocation problem between complementary observables that constrain different directions of the same inverse problem. Within the fixed-count model considered here, Fisher analysis provides a quantitative means of matching the detected-count allocation to the physical quantity or local uncertainty geometry that the measurement is intended to constrain.

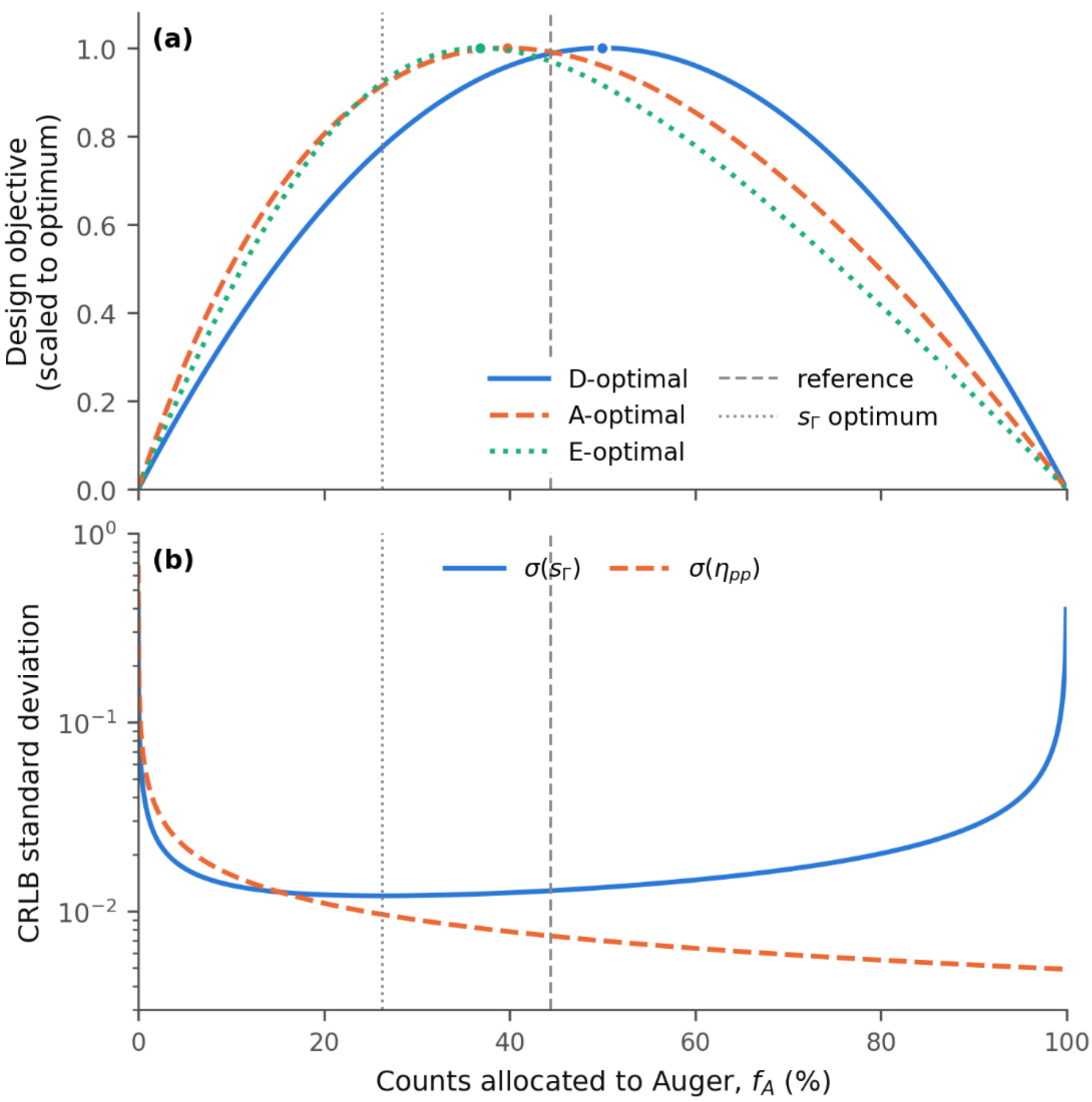

Figure 4: Information-optimal allocation of a fixed expected detected-count budget of $N_{tot} = 1.8 \times 10^5$ between XAS and final-state-resolved Auger spectroscopy. (a) Normalized D-, A-, and E-optimal design objectives as functions of the fraction $f_A$ allocated to Auger measurements. The corresponding optimal fractions are 0.5000, 0.3973, and 0.3684, respectively. (b) Marginalized Cramér–Rao lower-bound (CRLB) standard uncertainties of $s_\Gamma$ and $\eta_{pp}$ as functions of $f_A$, shown on a logarithmic vertical scale. The dotted vertical line indicates the $s_\Gamma$-specific optimum at $f_A = 0.2630$, while the dashed vertical line marks the reference allocation at $f_A = 0.4444$. The distinct optima demonstrate that the preferred detected-count allocation depends on the target parameter or multiparameter design criterion. These results concern expected detected counts rather than beam time or experimental cost.

## 3.5. Scope, Assumptions, and Limitations

The present analysis is designed to characterize parameter identifiability and information complementarity within a common state-resolved XAS–resonant-Auger model of $Ar^+$; it is not intended as a high-precision fit to a specific experiment. The quantitative Fisher results are therefore conditional on the adopted FAC atomic structure, spectral forward models, instrumental resolutions, count budgets, and two-parameter inference model.

The resonant-Auger treatment is an effective, incoherent, rate-based representation. The FAC rate tables used in this study provide state-resolved autoionization rates but do not contain the relative complex excitation−decay amplitudes required for a coherent Kramers−Heisenberg treatment. Interference between overlapping intermediate resonances is therefore not included. The adopted scan-integrated construction also assumes fixed excitation weights and a unit non-radiative yield in the normalized Auger branching model. Furthermore, the final-state-energy coordinate represents an effective spectral observable rather than a fully initial-state-resolved experimental electron binding-energy spectrum. Likewise, the finite-bandwidth study of Sec. 3.3 is implemented as effective Gaussian broadening with the adopted line-shape model and should not be interpreted as a coherent finite-bandwidth excitation calculation.

The calculated linewidths are conditional on the configuration spaces included in the FAC model. The radiative contribution to the intermediate-state widths contains only the explicitly calculated E1 channels, and a final $Ar^{2+}$ state assigned $\Gamma_f = 0$ is stable against secondary autoionization only within the included $Ar^{3+}$ configuration space. The simplified sequential-decay classification does not explicitly propagate the complete higher-charge-state cascade. The agreement of the calculated aggregate $Ar^{2+}/Ar^{3+}$ branching with available measurements therefore provides a consistency check on the overall decay partition rather than a level-by-level validation of the state-resolved final-state widths.

The statistical model is restricted to spectral-shape information under fixed expected detected-count normalization. Absolute yields, photon flux, and energy-dependent collection efficiency are not explicitly modeled; consequently, the allocation study of Sec. 3.4 concerns detected counts rather than beam time or experimental cost. Independent Poisson counting statistics are assumed. The adopted count budgets should be regarded as notional, particularly for the Auger observable: electron spectroscopy on ion beams is demanding because the target density of an ion beam is typically about six orders of magnitude lower than that of a neutral gas target [23]. The nuisance analysis is likewise selective: XAS-resolution uncertainty and a purely multiplicative gain are considered, whereas photon-energy calibration error, energy jitter, incident-spectrum uncertainty, structured backgrounds, detector nonlinearity, energy-dependent efficiency, and correlated noise are not modeled explicitly.

Finally, the inference is local in parameter space and conditional on the selected physical and statistical models. The Fisher matrix characterizes local distinguishability near $(s_\Gamma, \eta_{pp}) = (1, 1)$, for a parameterization consisting of a common primary-rate scale and a selective spectator-$pp$ rescaling. It does not establish global uniqueness of the inverse problem or test independent variations of additional decay groups. The finite-difference analysis establishes numerical stability of the local Fisher results, while the auxiliary Monte Carlo reconstruction tests the local CRLB approximation for the alternative spectral representation considered in Sec. 3.3.

# 4. Conclusion

This work examined the complementarity of X-ray absorption and resonant-Auger spectroscopy as an inverse problem in multichannel core-hole decay. Using a common state-resolved FAC description of the $Ar^+$ $2p \rightarrow 3d$ excitation–decay manifold, the two observables were compared through the Fisher-information geometry of a two-parameter model consisting of a common primary autoionization-rate scale, $s_\Gamma$, and a selective spectator-$pp$ rescaling, $\eta_{pp}$. The central question was not simply whether XAS and Auger produce spectroscopically distinct signals, but whether they constrain different physical combinations of the underlying decay dynamics.

The resulting information geometry is strongly anisotropic. XAS is sensitive to both parameters, but the corresponding XAS spectral derivatives are nearly collinear because the spectator-$pp$ fraction varies only weakly across the populated core-excited manifold. In the limiting case of an identical spectator-$pp$ fraction for all intermediate states, the two XAS derivatives become locally proportional and the XAS Fisher matrix becomes exactly rank one. The observed near-singularity is therefore a consequence of similar channel composition across the intermediate states rather than weak overall spectral sensitivity. By contrast, within the adopted normalized final-state-resolved Auger model, the observable is invariant under the common primary-rate scaling $s_\Gamma$, while remaining strongly sensitive to the channel-selective parameter $\eta_{pp}$. Auger therefore supplies complementary information that substantially reduces the channel-decomposition near-degeneracy left by XAS alone.

This distinction is reflected directly in the propagated physical uncertainties. Under the reference count and resolution model, XAS alone constrains the excitation-weighted total natural width to a local fractional CRLB of approximately 1.2%, and adding Auger reduces its standard uncertainty by only 0.70%. The improvement in channel-resolved quantities is much larger: the CRLB standard uncertainty is reduced by approximately 99.19% for the excitation-weighted spectator-$pp$ branching fraction, 97.82% for the non-$pp$ rate scale, and 96.25% for the $pp$ partial-rate scale. The generalized Fisher comparison gives one unchanged generalized-information mode and one strongly enhanced mode, $\lambda_1 = 1,\ \lambda_2 \simeq 1.5259 \times 10^4$, corresponding to an improvement factor of approximately $\sqrt{\lambda_2} \simeq 123.5$ on the associated standard-deviation scale. The principal contribution of Auger is therefore not a substantially improved determination of the total lifetime broadening, but the resolution of the decay-channel decomposition hidden within that width.

The robustness analyses support this interpretation while defining its numerical and model-dependent context. Finite-difference tests show that the numerical Fisher values and joint CRLBs are stable with respect to the derivative step, whereas the rank-one Auger structure itself follows analytically from the normalized branching model. An auxiliary kinetic-energy representation changes the Auger information by only a few percent and preserves the same degeneracy-breaking mechanism, while explicit Poisson reconstruction is consistent with the local Fisher prediction within the sampling precision of the Monte Carlo ensemble. Uncertainty in the calibrated XAS resolution primarily degrades inference of the absolute decay scale, whereas the Auger-driven determination of $\eta_{pp}$ remains essentially unchanged over the tested range.

The same anisotropy has a direct implication for measurement design. Under a fixed total budget of expected detected counts, the $s_\Gamma$-specific optimum assigns only about 26% of the events to Auger spectroscopy, reducing the marginalized uncertainty in $s_\Gamma$ by approximately 6.1% relative to the reference allocation. This result illustrates that the importance of complementary information does not imply count dominance: in the two-parameter problem, Auger need not receive the majority of counts to substantially reduce the channel-decomposition near-degeneracy that additional XAS counts alone resolve only inefficiently. Different D-, A-, and E- optimality criteria select different allocations because they optimize different aspects of the same anisotropic uncertainty geometry. These results concern expected detected-count allocation rather than beam time or experimental cost.

Within the scope defined in Sec. 3.5, the main conclusion is structural: XAS and Auger can be strongly complementary even when the second observable produces little improvement in the total linewidth-like quantity. In the present $Ar^+$ model, XAS primarily constrains the total decay-scale combination, whereas Auger supplies channel-selective information that enables a much more precise separation of the underlying partial decay contributions. The value of the combined measurement therefore lies in resolving a nearly degenerate inverse problem. More generally, the usefulness of a complementary spectroscopic observable is determined not only by how different its spectrum appears or how much it improves a single parameter, but by whether it adds an information direction that is missing or weakly constrained in the original measurement.

# Appendix A

## Monte Carlo Reconstruction Using the Auxiliary Auger Representation

Figure A1 presents the joint XAS−Auger parameter reconstructions obtained from 1000 independent Poisson realizations using the auxiliary kinetic-energy Auger representation described in Sec. 3.3. The reconstructed parameter pairs are distributed around the reference point $(s_\Gamma, \eta_{pp})$ = (1, 1), exhibiting a negative correlation consistent with the local Fisher-information prediction.

The empirical correlation coefficient was −0.397, compared with the corresponding Fisher prediction of −0.377. The Monte Carlo-to-CRLB standard-deviation ratios were 0.975 for $s_\Gamma$ and 0.997 for $\eta_{pp}$. These results support the local Fisher approximation within the tested auxiliary representation. They should not be interpreted as an independent validation of the final-state-resolved Auger model used for the principal information analysis.

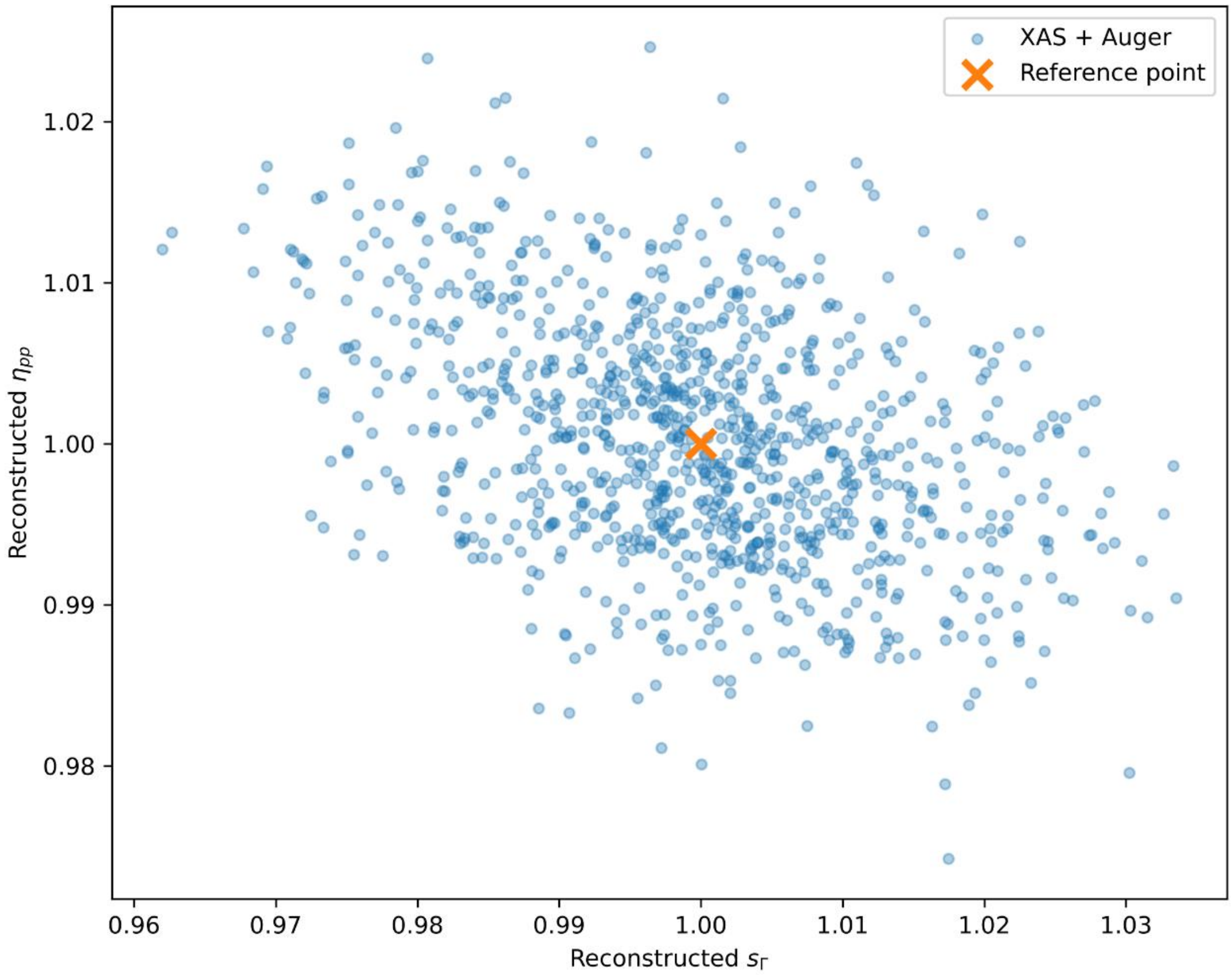

*Figure A1: Joint XAS–Auger Monte Carlo reconstruction obtained from $10^3$ independent Poisson realizations using the auxiliary Auger representation described in Sec. 3.3. The reconstructed parameter pairs cluster around the reference point ($s_\Gamma, \eta_{pp}$) = (1, 1) and exhibits the negative correlation discussed in the text.*